\documentclass[baaa]{baaa}

\usepackage[pdftex]{hyperref}
\usepackage{subfigure}
\usepackage{natbib}
\usepackage{helvet,soul}
\usepackage[font=small]{caption}

\contriblanguage{1}

\contribtype{1}

\thematicarea{7}

\begin{document}


\title{Disrupted and surviving satellites of Milky-Way-mass galaxies: connecting their properties with their host accretion histories}


\titlerunning{Disrupted an surviving MW satellites}


\author{
S.E. Grimozzi\inst{1,2},
M.E. De Rossi\inst{1,2},
\&
A.S. Font\inst{3}
}

\authorrunning{Grimozzi et al.}


\contact{salvy.eg279@gmail.com}


\institute{
Departamento de F{\'\i}sica, FCEN--UBA, Argentina \and   
Instituto de Astronomía y Física del Espacio, CONICET--UBA, Argentina
\and
Astrophysics Research Institute, Liverpool John Moores University, Reino Unido
}


\resumen{Existen dos poblaciones de galaxias enanas que pueden ser asociadas a la Vía Láctea (VL): las acretadas que se fusionaron completamente con ella en el pasado y las sobrevivientes que orbitan a su alrededor hoy en día. En este trabajo, analizamos la composición química del gas frío en ambas poblaciones asociadas a galaxias tipo VL en las simulaciones \texttt{ARTEMIS}. Encontramos que, a masa estelar fija, las acretadas tienen menor metalicidad y mayor enriquecimiento de elementos $\alpha$ que las sobrevivientes. A masa estelar fija, las acretadas con menor metalicidad y mayor $\mathrm{[Mg/Fe]}$ presentan mayores fracciones de gas capaz de formar estrellas y fueron acretadas a \textit{redshifts} más altos. En el caso de las sobrevivientes, encontramos una tendencia similar tanto para la metalicidad como para la abundancia de $\mathrm{[Mg/Fe]}$ en el gas.}

\abstract{Two populations of dwarf galaxies can be associated with the Milky Way (MW): the disrupted dwarfs that fully merged with it in the past and the surviving satellites that orbit around it in the present time. In this work, we analyzed the chemical composition of the cold gas in both populations of dwarfs associated with MW like galaxies in the \texttt{ARTEMIS} simulations. We found that, at fixed stellar mass, disrupted dwarfs have lower gas phase metallicity and are more $\alpha-$enhanced than surviving dwarfs. We also noticed that disrupted satellites accreted earlier and with higher SF gas fractions had lower metallicity and higher $\mathrm{[Mg/Fe]}$ at fixed mass. In the case of surviving dwarfs, we obtained a similar trend for both gas-phase metallicity and $\mathrm{[Mg/Fe]}$ abundance.}




\maketitle
\section{Introduction}\label{sec:intro}

According to the standard Lambda cold dark matter ($\Lambda\mathrm{CDM}$) model, the formation of galaxies similar to the Milky Way (MW) is a hierarchical process (\citealt{WhiteFrenk1991}). Large galaxies accrete and disrupt smaller structures like dwarf galaxies, thus increasing their masses and changing their chemical compositions. The study of the tidal debris generated by these accretion events can provide useful insights about the merger history of the MW, which could help us determine whether it is typical among other similar galaxies (\citealt{Naidu2020}). It has been found that these disrupted satellites exhibit different properties compared to the surviving ones that orbit the MW today (\citealt{Naidu2022}). Stellar populations in our galaxy that were found to be the remains of dwarf galaxies accreted long ago have chemical compositions different from those of present-day satellites. Similar differences have been found in MW-like systems in cosmological simulations (\citealt{Khoperskov2023},\citealt{Grimozzi2024}).

Since star formation is fueled by cold gas, the gas-phase metallicity ($Z_{\mathrm{gas}}$) of these dwarf galaxies affects the chemical composition of their stars. This implies that the different abundances of elements present in the stars of both disrupted and surviving dwarfs may be related to differences in their gas phases. For example, it has been observed that the mass-metallicity relation (MZR) of the gas is similar to that of stars, while both $Z_{\mathrm{gas}}$ and stellar metallicity ($Z_{*}$) increase with stellar mass ($M_{*}$) (\citealt{Tremonti2004}, \citealt{Gallazzi2005}, \citealt{Gallazzi2014}), although the former flattens at high masses (\citealt{Fraser-McKelvie2022}). Studies on the evolution of the MZR have also shown that metallicity decreases with redshift in both phases (for the gas phase, see \citealt{Tremonti2004}, \citealt{Andrews2013}, \citealt{Curti2020}, for the stellar phase, see \citealt{Gallazzi2005}, \citealt{Zahid2017}, \citealt{Cullen2019}).

In the case of disrupted dwarfs, it is difficult to trace the remnants of the tidally disrupted gas inside the MW. Here, cosmological simulations can play an important role, by providing more insights into the chemical composition of the gas in the accreted galaxies. In a previous work, we studied the stellar chemical properties of accreted dwarfs in the \texttt{ARTEMIS} simulations (\citealt{Grimozzi2024}). Here, we present complementary results on the properties of the gas phase of these dwarfs.

\section{Simulations and samples}

\subsection{\texttt{ARTEMIS} simulations}\label{subsec:ARTEMIS}

\texttt{ARTEMIS} is a collection of cosmological hydrodynamical simulations of Milky Way-mass halos (\citealt{Font2020}) performed using the \texttt{GADGET-3} code (\citealt{Springel2005}). The simulations adopt the same subgrid physical model as in the \texttt{EAGLE} simulations (\citealt{Schaye2015}), except a recalibration of parameters related to the efficiency of stellar feedback. Run in periodic cubic volumes of $25~\mathrm{Mpc}~h^\mathrm{{-1}}$ on a side, the ``zoomed-in" technique is applied, which allows higher spatial and particle resolutions around MW-mass like galaxies (\citealt{Bertschinger2001}). The initial masses for gas and dark matter particles in \texttt{ARTEMIS} are $m_{\mathrm{gas}}=2.23\times10^{4}~\mathrm{M_{\odot}}$ and $m_{\mathrm{DM}}=1.17\times10^{5}~\mathrm{M_{\odot}}$, respectively,  which are lower than those in the high resolution L025N0752 runs in \texttt{EAGLE}, with initial particle masses of $m_{\mathrm{gas}}=2.26\times10^{5}~\mathrm{M_{\odot}}$ and $m_{\mathrm{DM}}=1.21\times10^{6}~\mathrm{M_{\odot}}$. The main subgrid physical processes included in \texttt{ARTEMIS} are metal-dependent radiative cooling, hydrogen reionization, star formation, stellar feedback and evolution, black hole formation and evolution (through gas accretion and mergers), and feedback from active galactic nuclei.

\subsection{Definitions}

Throughout Sect. \ref{subsec:chemical_comp_and_acc_hist}, we study the merging history of MW-mass galaxies by using different proxies. One is the redshift of accretion ($z_{\mathrm{acc}}$), defined as the redshift when the dwarf reaches its maximum $M_{*}$. This choice is motivated by the finding that, as they cross the virial radius ($R_{200}$\footnote{Radius within which the density is 200 times the critical density of the Universe}) of their host, most disrupted dwarfs continue to form stars until they begin to disrupt. For surviving dwarfs, which typically accrete late, we use instead $z_{50}$, defined as the median redshift at which they reached $\sim50\%$ of their maximum stellar mass. The values of stellar mass ($M_{*}$), star-forming gas mass ($M_{\mathrm{SF}}$), and of the abundances of individual chemical elements used in this work correspond to the SubHalo catalogues in \texttt{ARTEMIS}, which are equivalent to those in EAGLE database (see \citealt{McAlpine2016})\footnote{These masses are calculated by adding the masses of all the particles assigned to an independent subhalo by the simulation rather than those within a specific radius. However, this selection doesn't affect the general trends in scaling relations involving those masses (see \citealt{DeRossi2017}; \citealt{LaraLopez2019}).}. 

\section{Results}\label{sec:results}

\begin{figure}[h]
\centering
\includegraphics[width=0.7\columnwidth]{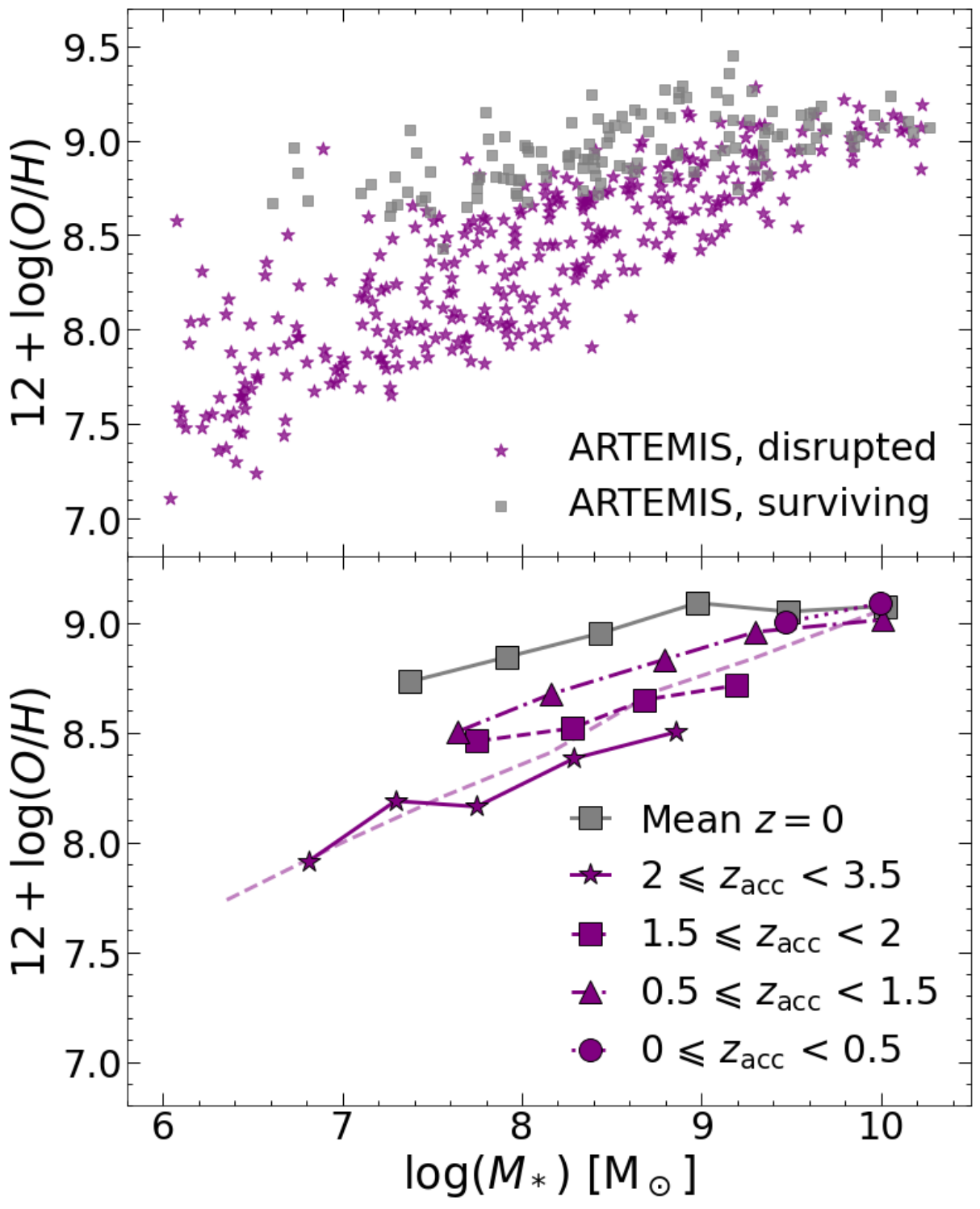}
\caption{\textit{Top}: The gas-phase MZR of disrupted (purple stars) and surviving (grey squares) dwarfs. Oxygen abundance is used as a proxy for metallicity. \textit{Bottom}: Mean MZR of surviving dwarfs at $z=0$ (grey) and of disrupted dwarfs at their accretion $z=z_{\mathrm{acc}}$ (purple), separated in four redshift bins. The purple dashed line in the background indicates the median of the whole disrupted population. Most surviving dwarfs are devoid of SF gas,  hence their low numbers.}
\label{fig:MZR}
\end{figure}

\subsection{Gas-phase chemical composition}


For star-forming gas, we use the oxygen abundance as a proxy for the gas metallicity. In the top panel of Fig.~\ref{fig:MZR}, we show the gas-phase MZR of all disrupted and surviving dwarfs in \texttt{ARTEMIS} containing at least one particle of star-forming gas ($m_{\mathrm{SF}} \geq 2.23\times10^{4}~\mathrm{M_{\odot}}$). In accordance with observations, there is a clear correlation between the oxygen abundances and stellar masses in both populations. At fixed stellar mass, surviving dwarfs are more chemically enriched than disrupted ones, with the exception of most massive galaxies ($M_{*}\gtrsim10^{9.5}~\mathrm{M_{\odot}}$). In addition, the scatter in oxygen abundances is larger for disrupted dwarfs and decreases with mass. The dwarfs that reached the highest masses  had more time to form stars and were therefore accreted at later times, whereas less massive ones could be accreted at different redshifts.

The scatter in metallicity is larger for disrupted dwarfs than for surviving ones. This can also be explained by the disrupted dwarfs having been accreted at different redshifts, unlike the surviving dwarfs (shown here at $z=0$). The evolution of the mean gas-phase MZR, shown in the bottom panel of Fig. \ref{fig:MZR}, supports this explanation. Here, the disrupted population is separated into four sub-populations based on their $z_{\mathrm{acc}}$. Galaxies that were accreted earlier had lower metallicity at accretion, whereas those that were accreted near $z=0$ have similar abundances to their surviving counterparts. A similar evolutionary trend is seen in observations (\citealt{Zahid2014}), as well as in the stellar debris of disrupted dwarfs in \texttt{ARTEMIS} (\citealt{Grimozzi2024}).

\begin{figure}[h]
\centering
\includegraphics[width=0.75\columnwidth]{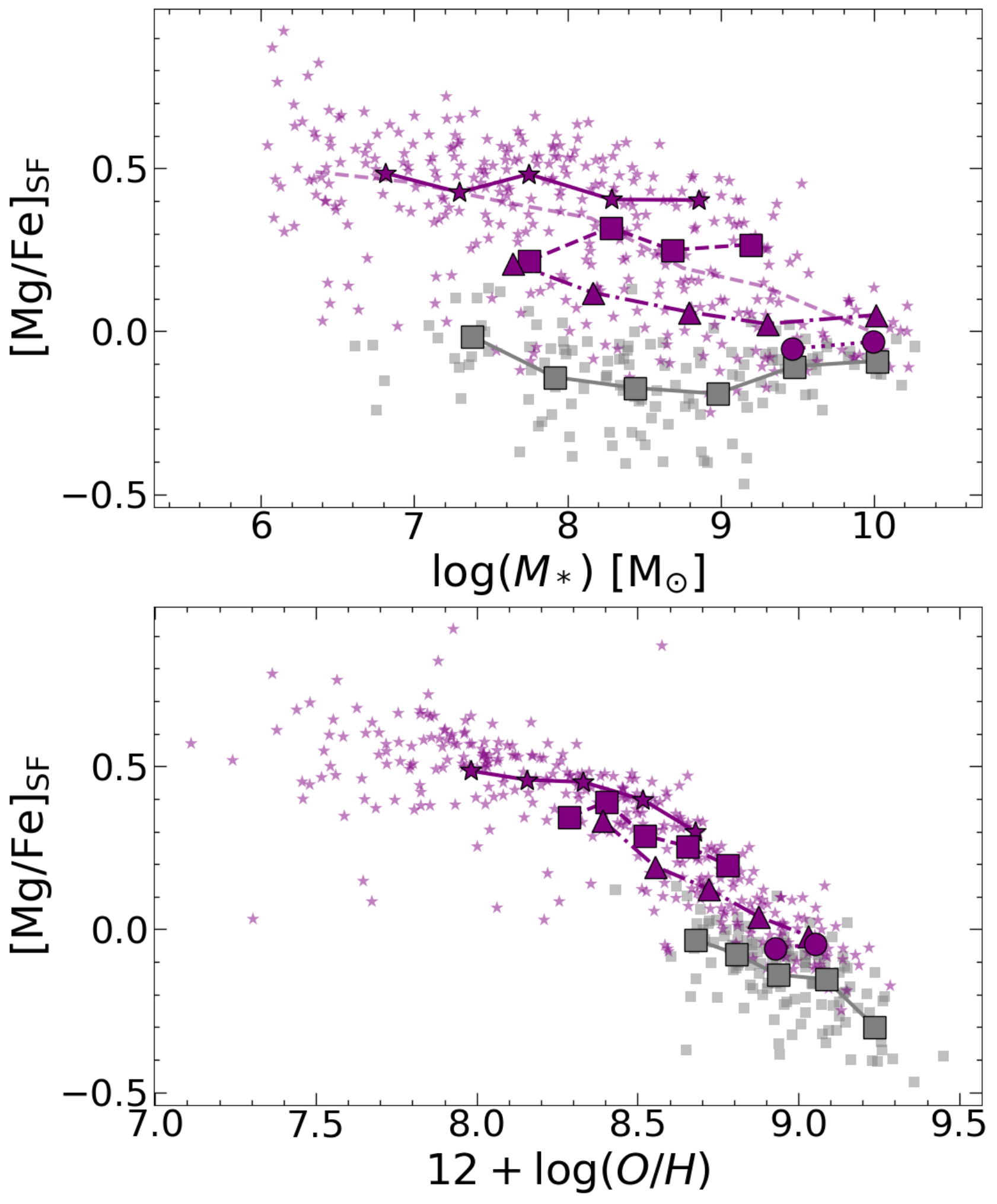}
\caption{Gas-phase $\mathrm{[Mg/Fe]}-M_{*}$ (\textit{top}) and $\mathrm{[Mg/Fe]}-$ metallicity (\textit{bottom}) relations for disrupted (purple stars) and surviving (grey squares) dwarf galaxies in the \texttt{ARTEMIS} simulations. See Fig. \ref{fig:MZR} for the legend.}
\label{fig:alpha_enhancement}
\end{figure}

The $\alpha-$enhancement is estimated by the relative abundance of magnesium over iron in the star-forming gas, $\mathrm{[Mg/Fe]_{\mathrm{SF}}}$. The top panel of Fig.~\ref{fig:alpha_enhancement} shows the $M_{*}-\mathrm{[Mg/Fe]_{\mathrm{SF}}}$ relationship of both disrupted dwarfs and surviving dwarfs. At fixed mass, the [Mg/Fe] in the SF gas is higher in disrupted dwarfs than in surviving ones. For disrupted dwarfs, [Mg/Fe]$_{\mathrm{SF}}$ also decreases as $M_{*}$ increases. Similar trends were found for the stellar component of disrupted galaxies in \texttt{ARTEMIS} (\citealt{Grimozzi2024}) and may be explained by the fact that early accreted galaxies tend to experience more bursty star formation processes than those accreted later (\citealt{Robertson2005}, \citealt{Khoperskov2023}). The fact that $\mathrm{[Mg/Fe]_{SF}}$ increases with $z_{\mathrm{acc}}$ supports this hypothesis.

The relation between metallicity and the $\alpha-$enhancement is shown in the bottom panel of Fig. \ref{fig:alpha_enhancement}. A similar trend is shared by both populations: the more oxygen rich the SF gas of a dwarf is, the lower its $\alpha-$element abundances are. This relation shows a dependence with $z_{\mathrm{acc}}$ as well. We also note that it has a knee, with a
steeper decline of $\mathrm{[Mg/Fe]}_{\mathrm{SF}}$ in galaxies at the high-metallicity end. This knee is located in the region delimited by $8.4\lesssim 12+\log(O/H)\lesssim 8.7$ and $0.25\lesssim \mathrm{[Mg/Fe]}_{\mathrm{SF}} \lesssim 0.50$, where the average $z_{\mathrm{acc}}$ is around $2$, close to the observed maximum of the Star Formation Rate density of the Universe (e.g., \citealt{Madau2014}).

\begin{figure*}
\centering
\includegraphics[width=1.0\textwidth]{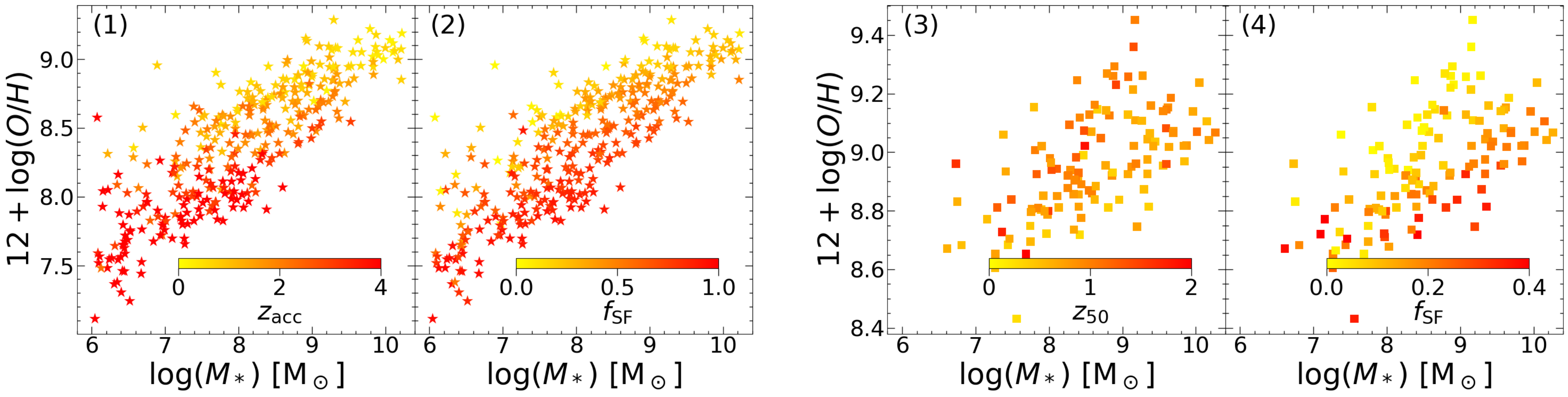}
\caption{\textit{Left panels}: gas-phase MZR of disrupted dwarfs color-coded by $z_{\mathrm{acc}}$ (1) and star forming gas fraction (2). \textit{Right panels}: gas-pase MZR of surviving dwarfs color-coded by $z_{\mathrm{50}}$ (3) and star forming gas fraction (4).}
\label{fig:MZR_color-coded}
\end{figure*}

\subsection{Chemical composition and accretion histories}\label{subsec:chemical_comp_and_acc_hist}

Figures \ref{fig:MZR} and \ref{fig:alpha_enhancement} show that the scatter in the gas-phase MZR and in the gas-phase $M_{*}-\mathrm{[Mg/Fe]_{\mathrm{SF}}}$ relation is larger for the disrupted population. Since disrupted dwarfs were accreted by their MW-mass hosts at various redshifts, the difference in the mentioned scatter may be caused by hosts experiencing different accretion histories. In order to study this effect, we evaluated the previous relations according to parameters associated with the accretion history of a galaxy. For disrupted dwarfs, we used the $z_{\mathrm{acc}}$ and the fraction of cold star-forming gas ($f_{\mathrm{SF}}$, which correlates with the star formation rate),  whereas for surviving dwarfs, instead of $z_{\mathrm{acc}}$, we evaluated $z_{50}$, i.e., when they were not yet accreted.

In Fig. \ref{fig:MZR_color-coded}, we color-code the gas-phase MZR of both populations with the aforementioned parameters. Panel (1) shows that there is a strong anti-correlation between $z_{\mathrm{acc}}$ and both SF gas metallicity and stellar mass; the more massive and oxygen-rich dwarfs were accreted at later times. Panel (2) shows the same trend for $f_{\mathrm{SF}}$. In the case of surviving dwarfs (right panels), there is no notable correlation between $z_{50}$ and either metallicity (Panel 3) or stellar mass (Panel 4). This can be attributed to the fact that galaxies without SF gas are not included in this sample, most of which have low masses and high $z_{\mathrm{50}}$ (see \citealt{Grimozzi2024}). For $f_{\mathrm{SF}}$, the correlation is weaker than in the case of disrupted satellites, but generally, oxygen-poor dwarfs have higher star-forming gas fractions.

\begin{figure*}
\centering
\includegraphics[width=1.0\textwidth]{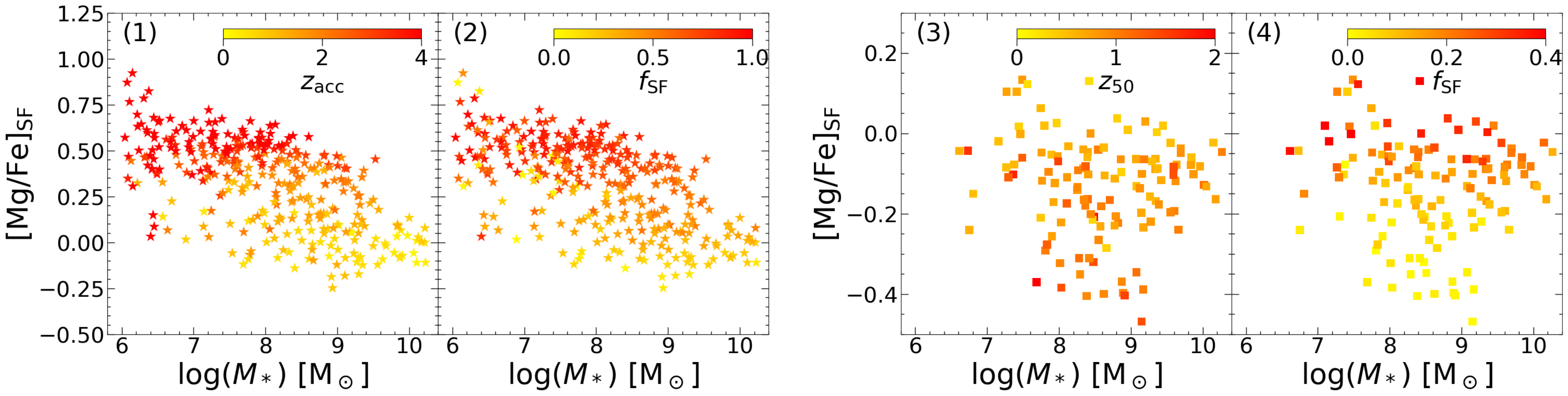}
\caption{\textit{Left panels}: gas-phase $\mathrm{[Mg/Fe]}-M_{*}$ relation of disrupted dwarfs color-coded by $z_{\mathrm{acc}}$ (1) and star forming gas fraction (2). \textit{Right panels}: gas-phase $\mathrm{[Mg/Fe]}-M_{*}$ relation of surviving dwarfs color-coded by $z_{\mathrm{50}}$ (3) and star forming gas fraction (4).}
\label{fig:alpha_enhancement_color-coded}
\end{figure*}

The $\mathrm{[Mg/Fe]_{\mathrm{SF}}}-M_{*}$ relation, color-coded as the MZR in Fig. \ref{fig:MZR_color-coded}, is shown in Fig. \ref{fig:alpha_enhancement_color-coded}. Panel (1) shows that the $\alpha-$enhancement of disrupted satellites correlates strongly with $z_{\mathrm{acc}}$: dwarfs with higher gas-phase $\mathrm{[Mg/Fe]}$ abundances were accreted at earlier times. As in the analysis of Fig. \ref{fig:alpha_enhancement}, this correlation can be associated with more bursty star formation histories of early accreted galaxies. Another possible factor is the delay time between the occurrence of Type II and Type Ia supernovae (SN~II and SN~Ia, respectively), with the former taking place in earlier than the latter. SNII produce more $\alpha-$elements and less iron than SNIa, which may contribute to the increase in $\mathrm{[Mg/Fe]}$ abundances in the early accreted dwarfs. Panel (2) shows that $f_{\mathrm{SF}}$ has a similar behavior, with the early accreted, more $\alpha-$enhanced dwarfs having larger SF gas reservoirs. We note the existence of a small number ($\lesssim10$) of SF-gas deprived disrupted dwarfs accreted at high redshift. The gas content of dwarf galaxies can be affected by the environment surrounding them, especially in the case of less massive satellites with low gravitational potentials. These structures may not be able to retain most of their gas reservoirs as they orbit around their more massive hosts after accretion, which could explain the existence of this small sub-population.

The right panels of Fig. \ref{fig:alpha_enhancement_color-coded} show the same relation for surviving dwarfs with $z_{50}$ instead of $z_{\mathrm{acc}}$. It is difficult to identify a clear trend in the case of the Panel 3, although it is interesting to note that surviving satellites with lower $z_{50}$ are not present among those with the lowest $\mathrm{[Mg/Fe]}$ abundances. Regarding the SF gas fraction presented in \ref{fig:alpha_enhancement_color-coded} (Panel 4), we observe the same trend as in the disrupted population: satellites with lower $\alpha-$enhancement contain less cold gas. These dwarfs may have been gravitationally bound to their main hosts for longer times, forming stars as they orbit them, decreasing their $\alpha-$element abundances and depleting their gas reservoirs with time.

\section{Conclusions}

We analyzed the properties of the gas-phase in both disrupted and surviving dwarfs in MW-mass hosts in the \texttt{ARTEMIS} cosmological simulations. This study was able to obtain trends for the star-forming gas which are similar to those found previously for the stellar components by \cite{Grimozzi2024}. The results of this work can be summarized as follows:

\begin{itemize}
    \item We obtained the gas-phase MZR and the $M_{*}-\mathrm{[Mg/Fe]_{\mathrm{SF}}}$ relation of both disrupted and surviving dwarfs. At fixed stellar mass, the SF gas in the surviving population is richer in oxygen and less $\alpha-$enhanced than in the disrupted one.

    \item The gas-phase metallicity of disrupted satellites decreases with the redshift of accretion. Since the dwarfs in this population were accreted at different redshifts ($z_{\mathrm{acc}}$), this would explain the larger scatter in the MZR of their gas compared to the scatter in the surviving ones.

    \item The $\mathrm{[Mg/Fe]_{\mathrm{SF}}}$ abundances decrease as the metallicity of the gas increases in both populations. In the case of disrupted dwarfs, this relation presents a knee around $12+\log(O/H)\approx 8.5$.

    \item In the case of disrupted satellites, the scatter in both their gas-phase MZR and their $M_{*}-\mathrm{[Mg/Fe]_{\mathrm{SF}}}$ relation correlates with their $z_{\mathrm{acc}}$ and their fraction of SF gas at accretion. For surviving dwarfs, the latter fraction also correlates with the scatter in both relations, but no notable correlation was found between $z_{50}$ and the scatter.
\end{itemize}

\begin{acknowledgement}
We thank the reviewer of this article for the constructive comments that helped to improve this manuscript. The authors of this paper thank the members of the \texttt{EAGLE} team for making their cosmological simulation code available for the \texttt{ARTEMIS} project and John Helly for constructing the merger trees for the MW-mass haloes studied here. 
The {\texttt ARTEMIS} project has received funding from the European Research Council (ERC) under the European Union's Horizon 2020 research and innovation programme (grant agreement No 769130).  We acknowledge support from {\it Agencia Nacional de Promoci\'on de la Investigaci\'on, el Desarrollo Tecnol\'ogico y la Innovaci\'on} (Agencia I+D+i, PICT-2021-GRF-TI-00290, Argentina).
\end{acknowledgement}


\bibliographystyle{baaa}
\small
\bibliography{bibliografia}

\end{document}